\documentclass[article,pre]{revtex4}%
\usepackage{amsfonts}
\usepackage{amsmath}
\usepackage{amssymb}
\usepackage{graphicx}%
\begin{document}
\title[Bistable Gradient Networks]{Bistable Gradient Networks in the Thermodynamic Limit}
\pacs{PACS number}
\pacs{PACS number}
\pacs{PACS number}
\author{Patrick N. McGraw and Michael Menzinger}
\affiliation{University of Toronto}
\date{\today}

\begin{abstract}
We examine the large-network, low-loading behaviour of an attractor neural
network , the so-called bistable gradient network (BGN). \ We use analytical
and numerical methods to characterize the attractor states of the network and
their basins of attraction. \ The energy landscape is more complex than that
of the Hopfield network and depends on the strength of the coupling among
units. \ At weak coupling, the BGN acts as a highly selective associative
memory; \ the input must be close to the one of the stored patterns in order
to be recognized. \ A category of spurious attractors occurs which is not
present in the Hopfield network. \ Stronger coupling results in a transition
to a more Hopfield-like regime with large basins of attraction. The basins of
attraction for spurious attractors are noticeably suppressed compared to the
Hopfield case, even though the Hebbian synaptic structure is the same and
there is no stochastic noise. \ 

\end{abstract}
\maketitle

\section{ \bigskip Introduction}

Many neural network models\cite{Haykin}\cite{Amitbook}, \ in addition to their
potential applications to computation, robotics and artificial intelligence,
constitute intriguing dynamical systems in their own right, showing unusual
manifestations of the statistical mechanics phenomena of order, disorder and
frustration. \ The connection between neural networks and statistical
mechanics became especially clear with the introduction of the Hopfield
\cite{Hopfield}\cite{Little} model, \ which furnishes a model of associative
memory, or the recall of a memorized pattern from an incomplete stimulus.
\ This model has a well-defined energy function and is closely related to the
Sherrington-Kirkpatrick spin glass model \cite{SKModel}\cite{Mezard}.

In this paper we consider a Hopfield-like network of $\ N$ bistable elements,
the bistable gradient network or BGN, previously introduced in \cite{Chinarov}%
. \ A closely related model was also discussed in \cite{Hoppensteadt} and
suggested as a model for the so-called "bistability of perception" in the
interpretation of ambiguous visual stimuli \cite{Bistability}. \ The network's
dynamics consists of a continuous gradient descent described by the coupled
differential equations
\begin{equation}
\frac{dx_{i}}{dt}=-\frac{\partial H}{\partial x_{i}}, \label{gradienteqn}%
\end{equation}
where $x_{i}\;(1\leq i\leq N)$ are continuous-valued real state variables
associated with the $N$ nodes of the network and the Hamiltonian or energy
function is given by
\begin{align}
H  &  =H_{0}+H_{int}+H_{ext}\nonumber\\
&  =\sum_{i=1}^{N}\left(  \frac{x_{i}^{4}}{4}-\frac{x_{i}^{2}}{2}\right)
-\frac{1}{2}\gamma\sum_{i,j=1}^{N}w_{ij}x_{i}x_{j}-\sum_{i=1}^{N}b_{i}x_{i}.
\label{Hamiltonian}%
\end{align}
The quantities $w_{ij}$ are a symmetric matrix of coupling strengths, and the
quantities $b_{i}$ are bias terms or external inputs to the network. For the
remainder of this paper we will set all $b_{i}=0$ unless otherwise stated;
\ we include them here only for the sake of generality. $\gamma$ is a control
parameter determining the strength of the internode couplings relative to the
local terms. \ The variables $x_{i}$ can be viewed as components of an
$N$-dimensional state vector $\mathbf{x}$. \ We define a normalized inner
product between two state vectors $\mathbf{x}^{1}$ and $\mathbf{x}^{2}$ by
$\mathbf{x}^{1}\cdot\mathbf{x}^{2}\equiv\frac{1}{N}\sum_{i=1}^{N}x_{i}%
^{1}x_{i}^{2}$. \ The first term in the Hamiltonian represents a local
double-well potential for each node, making each node individually bistable.
\ This local potential constitutes the main difference between the BGN and the
Hopfield model. \ \ The classical Hopfield network (HN) which we consider by
way of comparison is described by the Hamiltonian
\begin{equation}
H_{HN}=-\frac{1}{2}\sum_{i,j=1}^{N}w_{ij}x_{i}x_{j}+\sum_{i=1}^{N}b_{i}x_{i}
\label{HopHam}%
\end{equation}
where the $x_{i}$ are now \emph{discrete} state variables restricted to the
values $\pm1$. \ Although continuous versions of \ the HN have also been
studied, \ these generally lack the bistability property, and their behaviour
is essentially similar to that of the discrete version \cite{ContinuousHop}.

The variables $x_{i}$ can be thought of as the outputs of proccessing units or
neurons. \ Their dynamical equations can be written as
\begin{equation}
\frac{dx_{i}}{dt}=x_{i}-x_{i}^{3}+h_{i}, \label{dynamiceqn}%
\end{equation}
where $h_{i}\equiv\gamma\sum_{j=1}^{N}w_{ij}x_{i}+b_{i}$ is the input to the
neuron from internal and external connections. By analogy with Ising spin
systems, we also refer to $h_{i}$ as a \textquotedblleft magnetic
field.\textquotedblright\ \ The steady-state output for a given input is a
solution of the fixed-point equation \
\begin{equation}
x_{i}-x_{i}^{3}+h_{i}=0. \label{fixedpt}%
\end{equation}
When $h=0$, there are stable fixed points at $x=\pm1$ and an unstable fixed
point at $x=0.$ \ An applied field shifts the positions of the fixed points.
\ A saddle-node bifurcation occurs when $\left\vert h\right\vert =h_{c}%
=\frac{2\sqrt{3}}{9}\approx0.385$ so that for larger values of the field there
is only one equilibrium, aligned with the field ($x$ and $h$ have the same
sign). \ $x$ is in principle unbounded; \ the output does not truly saturate
when the input is large. \ \ The double-valuedness and the lack of saturation
are the principal differences between the input-output relation for the BGN
and that of the Hopfield model, including its continuous\ versions. \ 

\ \ Numerous experimental studies have been made on intriguing chemical
analogs of the BGN (see, e.g. \cite{ChemNetE}, \cite{ChemNetM}). These studies
involved networks of bistable chemical reactors coupled either electrically or
through mass exchange. \ Previous computational work on small BGN's
\cite{Chinarov} \ suggested that under some conditions the network might
permit the storage of a larger number of patterns than in a HN of the same
size, without any modification of the basic Hebb learning rule. \ It was
noted, however, that the stability of a particular attractor can depend on the
control parameter $\gamma$. \ \ Some dependence of pattern stability on the
coupling strength had also been noted in the experiments on the mass-coupled
chemical networks \cite{ChemNetM}. \ \ 

In this paper we focus on the behaviour of the network in the case where the
number of nodes is large and the number of memorized patterns is small.
\ \ \ Using both analytical techniques and numerical
simulations\footnote{Simulations were conducted by integrating the
differential equations numerically using a fourth-order Runge-Kutta algorithm
with adaptive step size. \ The system was judged to have converged to a fixed
point if the magnitudes of all derivatives $\left\vert \partial H/\partial
x_{i}\right\vert $ fell below a convergence criterion which for most examples
was taken as 0.001. \ \ Our HN simulations, which were used for comparisons,
were performed using asynchronous updating in random order. \ Unless otherwise
indicated, all simulations were done on an $N=1000$ network with $p=5$ stored
memory patterns.}, we examine the retrieval of stored patterns and classify
the attractors that occur. \ We find that there are three types of attractors.
\ In addition to \emph{memory} or \emph{retrieval states}, there are spurious
attractors in which no pattern is fully recognized. \ These include the
\emph{mixture} or \emph{spin glass} \emph{states} familiar from HN studies, as
well as an additional category specific to the BGN which we refer to as
\emph{uncondensed states}. \ \ We examine how the attractors and their basins
of attraction change as the control parameter $\gamma$ is changed.
\ Throughout the paper, we compare our model to the zero-temperature or
deterministic discrete Hopfield model. \ \ It is hoped that these results can
illuminate some of the novel behavior of the BGN and clarify its relation to
the HN. \ \ The behavior of the BGN under higher memory loading and the
question of its maximum storage capacity will be addressed elsewhere. \ 

\section{Storage and retrieval of binary patterns}

As in previous work on Hopfield networks\cite{Hopfield}\cite{Amit1}%
\cite{Amitbook}, we define the task of associative memory as follows. \ We are
given a set of \ $p$ distinct $N$-dimensional vectors or \emph{memory patterns
}\ $\mathbf{\xi}^{\mu}$ $(\mu\in\{1,...p\}),$ which are to be recognized by
the network. \ \ The patterns should correspond to attractors of the network's
dynamics. \ We will refer to these attractors as \emph{retrieval states.
}\ Input is given by imposing a particular initial condition on the network.
\ If that initial condition is sufficiently close to one of the memorized
patterns, then the network's state should converge to the correct nearby
attractor, and we say that the network has recognized or retrieved the
pattern. \ In this paper we follow the HN literature in considering the case
where the patterns are random and uncorrelated strings of $+1$'s and $-1$'s.
\ We read the output of the network according to the \emph{signs} of the
$x_{i}$. \ Thus we say that the network has recalled pattern 1, for example,
if $\operatorname{sign}(x_{i})=\xi_{i}^{1}$ for all $i$. \ Although variations
in the magnitude of $x_{i}$ can be important to the dynamics, \ we will for
the moment ignore them for the purpose of reading the output. \ As we will see
below, the retrieval states in general do not have $\left\vert x_{i}%
\right\vert =1$ even though the patterns have $\left\vert \xi_{i}\right\vert
=1.$ \ We focus here on the limiting case $N\rightarrow\infty$, $p/N\ll1$,
\ or large networks with low memory loading. \ (Strictly speaking, we take $N$
to infinity while $p$ remains finite.) \ In this case the inner product of a
pair of patterns\textbf{\ }$\mathbf{\xi}^{\mu}\cdot\mathbf{\xi}^{\nu}=\frac
{1}{N}\sum_{i}\xi_{i}^{\mu}\xi_{i}^{\nu}$ behaves as a Gaussian random
variable with zero mean and variance $\frac{1}{\sqrt{N}}$, \ so that in the
$N\rightarrow\infty$ limit the pattern vectors are nearly orthogonal to each
other \ and form a basis for a $p$-dimensional subspace of the $N$-dimensional
configuration space.

As in the HN, we construct the coupling matrix from the stored patterns
according to the Hebb \cite{Hebb}learning rule: \
\begin{equation}
w_{ij}=\frac{1}{N}\sum_{\mu=1}^{p}\xi_{i}^{\mu}\xi_{j}^{\mu}-\frac{p}{N}%
\delta_{ij}. \label{Hebb}%
\end{equation}
The term $-(p/N)\delta_{ij}$ is included to make all diagonal elements of the
coupling matrix zero. Non-zero diagonal entries would have the effect of
adding an additional quadratic self-interaction term. \ Following the usual
practice we omit them here so that the quadratic term is contained only in the
local potential. \ \ For the case $p/N\rightarrow0$, however, the effect of
the diagonal elements is negligible and we can substitute the simpler learning
rule
\begin{equation}
w_{ij}=\frac{1}{N}\sum_{\mu=1}^{p}\xi_{i}^{\mu}\xi_{j}^{\mu}.
\label{simpleHebb}%
\end{equation}

A useful set of order parameters are the overlaps $m_{\mu}$, which are inner
products of the network's state with each of the stored patterns: \
\begin{equation}
m_{\mu}\equiv\mathbf{\xi}^{\mu}\cdot\mathbf{x}=\frac{1}{N}\sum_{i=0}^{N}%
\xi_{i}^{\mu}x_{i} \label{mdef}%
\end{equation}
For the discrete HN, these variables take values $-1\leq m_{\mu}\leq1$,
\ while for the BGN any real values are possible. \ It will be useful to
define another set of variables, which we will call ``bit overlaps,''
by\footnote{The bit overlaps $b_{\mu}$ should not be confused with the biases
or external fields $b_{i}$ in (\ref{Hamiltonian}), which are set to zero for
the remainder of this paper. \ }%
\begin{equation}
b_{\mu}\equiv\frac{1}{N}\sum_{i=0}^{N}\xi_{i}^{\mu}\operatorname{sign}(x_{i}).
\label{bdef}%
\end{equation}
\ The bit overlap is simply related to the Hamming distance by \ $b_{\mu
}=\frac{1}{2N}(N-2d(\mathbf{x},\mathbf{\xi}^{\mu}))$ where the Hamming
distance $d(\mathbf{x,y})$ between two vectors is defined as the number of
elements for which their signs differ, or the number of positions $i$ such
that $x_{i}y_{i}<0$. \ Unlike $m_{\mu},$ the bit overlaps always obey $-1\leq
b_{\mu}\leq1$. \ They encode information about sign agreements but not about
magnitudes of the outputs $x_{i}$. \ \ 

The definitions of the overlap variables allow us to rewrite the Hamiltonian
and the dynamical equations in useful forms. \cite{Amitbook} \ In particular,
if the synaptic matrix $\mathbf{w}$ is given by the simplified Hebb rule
(\ref{simpleHebb}), then the interaction term of the Hamiltonian can be
rewritten in terms of $m_{\mu}$ as follows: \
\begin{equation}
H_{int}=-\frac{\gamma}{2}\sum_{i,j=1}^{N}w_{ij}x_{i}x_{j}=-\frac{\gamma}%
{2N}\sum_{i,j=1}^{N}\sum_{\mu=1}^{p}x_{i}\xi_{i}^{\mu}\xi_{j}^{\mu}%
x_{j}=-N\frac{\gamma}{2}\sum_{\mu=1}^{p}(m_{\mu})^{2}, \label{mHam}%
\end{equation}
and the net input to a given node from the other nodes is given by
\begin{equation}
h_{i}\equiv-\frac{\partial H_{int}}{\partial x_{i}}=\gamma\sum_{j=1}^{N}%
w_{ij}x_{j}=\frac{\gamma}{N}\sum_{j=1}^{N}\sum_{\mu=1}^{p}\xi_{i}^{\mu}\xi
_{j}^{\mu}x_{j}=\gamma\sum_{\mu=1}^{p}\xi_{i}^{\mu}m_{\mu}. \label{mField}%
\end{equation}

\subsection{Retrieval states at low memory loading}

To show that the network functions properly as an associative memory, we
exhibit the attractor states corresponding to the stored patterns, demonstrate
their stability, and show that a pattern can be retrieved from an initial
condition which lies close to the pattern but differs from it by one or more
incorrect signs.

Consider the state $\mathbf{x}=M\mathbf{\xi}^{\nu}$ where $M$ is a scalar and
$\mathbf{\xi}^{\nu}$ is a particular one of the stored patterns. \ We will
show that for a suitable value of $M$ this state represents a stable fixed
point of the dynamics and is therefore the retrieval state we seek. \ In this
state, $m_{\nu}=M$, $b_{\nu}=1$, and all other overlaps are small. \ The field
acting on the $i$-th node can be written as follows:%
\begin{equation}
h_{i}=\gamma\xi_{i}^{\nu}M+\gamma\sum\limits_{\mu\neq\nu}\xi_{i}^{\mu}m^{\mu}.
\end{equation}
The sum over patterns $\mu\neq\nu$ is called the crosstalk term. For the
overlaps with these other patterns we have%
\begin{equation}
m^{\mu}=\frac{1}{N}\sum\limits_{i}x_{i}\xi_{i}^{\mu}=M\frac{1}{N}%
\sum\limits_{i}\xi_{i}^{\nu}\xi_{i}^{\mu}\qquad(\mu\neq\nu).
\end{equation}
Since the patterns are random, each of these overlaps is of order $M/\sqrt{N}%
$. The number of patterns remains finite as $N\rightarrow\infty$ , so the sum
in the crosstalk term vanishes in this limit and $\ h_{i}\approx\gamma\xi
_{i}^{1}M$ . \ A stationary state must satisfy the fixed point condition
(\ref{fixedpt}) for each node, which leads to a self-consistency condition on
$M$:
\begin{align}
0  &  =x_{i}^{3}-x_{i}-h_{i}=(M\xi_{i}^{\nu})^{3}-(1+\gamma)(M\xi_{i}^{\nu
})\nonumber\\
0  &  =M^{3}-(1+\gamma)M.
\end{align}
In the last step, we have used the fact that $\xi_{i}^{3}=\xi_{i}$ when
$\xi_{i}=\pm1$ and then divided out the common factor of $\xi_{i}$. Solutions
to this condition are an unstable equilibrium $M=0$ and two stable equilibria%
\begin{equation}
M=m_{\nu}=\pm\sqrt{1+\gamma}. \label{retrievalM}%
\end{equation}
The two stable solutions represent perfect retrieval of pattern $\nu$ and its
mirror state, respectively. \ The doubled state is a consequence of the
$Z_{2}$ symmetry in the Hamiltonian. \ Since the overlap $m_{\nu}$ is equal to
$M$ and all other overlaps vanish in the thermodynamic limit, the energy of
this retrieval state is easily calculated using expression (\ref{mHam}) for
the energy in terms of the overlap variables, giving:%
\begin{align}
E  &  =\sum_{i}\left(  \frac{x_{i}^{4}}{4}-\frac{x_{i}^{2}}{2}\right)
-\frac{\gamma}{2}\sum_{\mu}m_{\mu}^{2}\nonumber\\
&  =N\left(  \frac{M^{4}}{4}-\frac{M^{2}}{2}\right)  -\frac{\gamma}{2}%
M^{2}\nonumber\\
&  =N\left(  -\frac{1}{4}-\frac{\gamma}{4}-\frac{\gamma^{2}}{4}\right)
\label{retrievalE}%
\end{align}
Note that this energy expression is extensive (proportional to $N$) and a
monotonically decreasing function of $\gamma$. \ 

Having identified the state
\begin{equation}
\mathbf{x}^{\nu}=\sqrt{1+\gamma}\mathbf{\xi}^{\nu} \label{retrievalstate}%
\end{equation}
as an equilibrium state, we now demonstrate its stability using a linear
stability analysis of the dynamical equations%
\begin{equation}
\frac{dx_{i}}{dt}=y_{i}\equiv x_{i}-x_{i}^{3}+\gamma\sum\limits_{j}w_{ij}%
x_{j}.
\end{equation}
Evaluating the Jacobian $\partial y_{i}/\partial x_{j}$ at the fixed point
(\ref{retrievalstate}) we obtain: \ \ %

\begin{align}
\frac{\partial y_{i}}{\partial x_{j}}  &  =(1-3x_{i}^{2})\delta_{ij}+\gamma
w_{ij}=\left(  1-3(\gamma+1)\right)  \delta_{ij}+\gamma w_{ij}\nonumber\\
&  =(-2-3\gamma)\delta_{ij}+\gamma w_{ij}, \label{Jacobian}%
\end{align}
where $\delta_{ij}$ is the Kronecker delta. \ The fixed point is linearly
stable if and only if the Jacobian has no positive eigenvalues. \ This depends
in turn on the eigenvalues of the synaptic connection matrix $\mathbf{w}$.
\ But in the limit where all of the stored patterns $\mathbf{\xi}^{\mu}$ are
mutually orthogonal, the stored patterns are themselves eigenvectors spanning
a degenerate subspace with eigenvalue 1, while the complement of this subspace
has eigenvalue 0. \ (The Hebb rule (\ref{simpleHebb}) itself gives a spectral
decomposition of $\mathbf{w}$.) \ Since the maximum eigenvalue of $\mathbf{w}$
is 1, \ we see that the Jacobian at the retrieval fixed point has no positive
eigenvalues and so the retrieval state is linearly stable for any value of
$\gamma$. \ In fact, all eigenvalues become more negative as $\gamma$
increases.\ $\ $ We reiterate that this result is valid in the ideal limit of
large $N$ and low loading where the stored patterns are orthogonal. \ For
finite-sized networks with finite overlaps among the patterns, \ it is
possible for the memory states to be destabilized by the crosstalk terms.
\ This issue will be examined elsewhere. \ 

Numerical results for a network with $N=1000$ nodes and $p=5$ random patterns
agree excellently with the above description. \ To study a retrieval state
numerically, we initialized the network to the state $\mathbf{x}=\mathbf{\xi
}^{1}$ (arbitrarily choosing the first pattern). \ Starting at $\gamma=0$,
\ we increased $\gamma$ by small steps to $\gamma=6$. \ At each step, we
integrated the dynamical equations until they converged. \ This procedure
allows us to examine the evolution of a state under quasistatic changes in the
control parameter $\gamma.$ \ We verified that $b_{1}$ remained equal to 1
over the whole range $0\leq\gamma\leq6,$ indicating that the retrieval state
is stable. \ The measured values of $m_{1}$ and $E$ were within 1\% of the
theoretical expressions (\ref{retrievalM}) and (\ref{retrievalE}) respectively.

\subsection{Error correction and basins of attraction\label{errorcorrect}}

Linear stability analysis has shown that the retrieval states are stable
against \emph{infinitesimal} perturbations for any value of $\gamma$, but this
does not guarantee their stability against the flipping of signs of one or
more nodes. \ In order to function as an associative memory, a network must be
capable of dynamically correcting sign errors. \ When presented with an input
at a small, nonzero Hamming distance from one of the stored patterns (i.e.,
differing from it by a few reversed signs) it must be able to flip the
reversed signs and restore the correct pattern. \ We will now show that there
is a critical value $\gamma_{c}=\frac{1}{3}$ above which the correction of
sign-flip errors can occur. \ For smaller values of $\gamma$ the BGN does not
correct sign errors and thus does not truly function as an associative memory,
but as $\gamma$ increases above $\frac{1}{3}$ the retrieval states develop
increasingly large basins of attraction. \ 

Consider a state of the network which is a slightly corrupted retrieval state:
\ all node variables have the values $x_{i}=M\xi_{i}^{\nu}=\sqrt{1+\gamma}%
\xi_{i}^{\nu}$ with the exception of one or possibly some number $\ll N$ of
nodes which may be misaligned. \ In such a state the few misaligned nodes make
only a small contribution to the overlap sums, so we have $m_{\nu}\approx M$
and $m_{\mu}\approx0(\mu\neq\nu)$. \ The field acting on each node is
therefore $h_{i}\approx\gamma M\xi_{i}^{\nu}.$ The misaligned bits experience
a field opposite to their signs. \ If the field becomes larger than the
critical value $2\sqrt{3}/9,$ then there is only one stable equilibrium for
each node, and the misaligned nodes will flip to conform with the stored
pattern. \ Error correction therefore occurs if%

\begin{equation}
\left\vert h_{i}\right\vert \approx\gamma M\approx\gamma\sqrt{1+\gamma}%
>\frac{2\sqrt{3}}{9}\approx0.385.
\end{equation}
The critical value occurs when the equality $\gamma\sqrt{1+\gamma}%
=\frac{2\sqrt{3}}{9}$ holds, or at $\gamma_{c}=\frac{1}{3}$. \ \ 

If the pattern is more strongly corrupted (a significant number of bits
misaligned) then the situation is more complicated, because the presence of a
larger number of misaligned bits may reduce the value of $m_{\nu}$ and thus
the magnitude of the field. \ The misaligned bits have a significant
back-reaction on the ones with the correct sign. \ The correction of larger
numbers of sign errors requires higher values of $\gamma$. \ We will return to
this point later; the basic result is that when $\gamma$ is only slightly
above the threshold of $\frac{1}{3},$ the memory states have rather small
basins of attraction, but these basins grow as $\gamma$ increases. \ 

\section{Spurious attractors: Spin glass states\label{mixture}}

In the case of the HN, the Hebb learning rule results in a large number of
``spurious'' attractors in addition to the retrieval states. \ The energy
function defines a rugged landscape, and a trajectory which does not start
sufficiently close to one of the stored patterns may become trapped in one of
the spurious local minima instead of one corresponding to a recalled pattern.
\ It is possible to suppress the spurious minima by introducing thermal noise
which allows trajectories to jump out of the shallower basins of attraction
into deeper ones. \ 

At low levels of loading, the HN possesses spurious attractors which are
nonlinear combinations of the stored patterns. There is a hierarchy of
symmetric \emph{mixture states} of the form\cite{Amit1}\cite{Amitbook}%
\begin{equation}
x_{i}=\operatorname{sign}(\xi_{i}^{\mu_{1}}\pm\xi_{i}^{\mu_{2}}\pm...\pm
\xi_{i}^{\mu_{n}})
\end{equation}
These states overlap equally with $n$ different patterns: \ $m_{\mu_{1}}%
^{2}=m_{\mu_{2}}^{2}=...m_{\mu_{n}}^{2}<1$. \ \ For the HN, only the mixtures
with odd $n$ are stable. \ The $n=3$ mixtures have the lowest energy in this
category, and the energies increase with $n,$ asymptotically approaching
$-1/\pi$. \ As the number of stored patterns $p$ increases, these spurious
states proliferate exponentially; their number is of order $3^{p}$. \ There
are also non-symmetric mixtures. The proliferation of spurious states is
associated with spin glass type behaviour in the HN. \ Accordingly we also
refer to these mixed states somewhat loosely as \emph{spin glass states}.
\ \footnote{More precisely, the mixture states \textquotedblleft
melt\textquotedblright\ into spin glass states when the number of memorized
patterns becomes higher. \ With higher numbers of patterns, there is a
distinction between mixture states which overlap with a small set of the
memory patterns and spin glass states which overlap with nearly all, but that
distinction is not salient for the cases we consider with very few patterns.
\ (See reference \cite{Amit1}.)}

We will show here that the BGN possesses mixture states analogous to those of
the HN, but their structure is slightly more complex. \ Let us focus on the
$n=3$ symmetric mixture state with positive signs. \ For the HN, this state is
given by
\begin{equation}
x_{i}=\xi_{i}^{S}\equiv\operatorname{sign}(\xi_{i}^{\mu_{1}}+\xi_{i}^{\mu_{2}%
}+\xi_{i}^{\mu_{3}}).
\end{equation}
This state is stable against individual sign flips because each node is
subject to a non-zero magnetic field which maintains its alignment. \ To see
this, note that there are two possibilities for each bit. \ Either all three
patterns agree at that particular site ($\xi_{i}^{\mu_{1}}=\xi_{i}^{\mu_{2}%
}=\xi_{i}^{\mu_{3}})$ or one of the patterns has the opposite sign from the
other two, for example $\xi_{i}^{\mu_{1}}=\xi_{i}^{\mu_{2}}=-\xi_{i}^{\mu_{3}%
}.$ \ When all three agree, we say that the $i$-th bit is a \textquotedblleft
unamimous\textquotedblright\ bit. \ If the patterns are random, then each
$\xi_{i}^{\mu}$ is $\pm1$ with equal probability, \ giving a probability of
$\frac{1}{4}$ that a given bit is unanimous. \ The mixture state has equal
overlaps with all three of the patterns. \ Since for a given $i$ there is a
probability of $\frac{3}{4}$ that $\xi_{i}^{S}=\xi_{i}^{\mu_{1}}$, we have
$m_{\mu_{1}}=\frac{1}{N}\sum_{i=1}^{N}\xi_{i}^{S}\xi_{i}^{\mu_{1}}=\frac{3}%
{4}-\frac{1}{4}=\frac{1}{2}$, and likewise $m_{\mu_{2}}=m_{\mu_{3}}=\frac
{1}{2}.$ \ \ The field acting at the $i$-th site is
\begin{equation}
h_{i}=\sum_{\mu}m_{\mu}\xi_{i}^{\mu}=\frac{1}{2}(\xi_{i}^{\mu_{1}}+\xi
_{i}^{\mu_{2}}+\xi_{i}^{\mu_{3}}).
\end{equation}
(Consistent with the low-loading, large-$N$ limit, we ignore all other
overlaps which are of order $\frac{1}{\sqrt{N}}$.) \ This gives $h_{i}%
=\frac{3}{2}\xi_{i}^{S}$ for unanimous bits and $h_{i}=\frac{1}{2}\xi_{i}^{S}$
for the others. \ In any case, each node of the HN experiences a field which
stabilizes its alignment. \ \ 

We will now show that the BGN, like the HN, possesses a mixture state in which
the sign of $x_{i}$ is given by the majority vote of three of the stored
patterns: \
\begin{equation}
\operatorname{sign}x_{i}^{S}=\operatorname{sign}(\xi_{i}^{\mu_{1}}+\xi
_{i}^{\mu_{2}}+\xi_{i}^{\mu_{3}})
\end{equation}
This state has a more complicated structure, however, because the magnitude of
$x_{i}$ at a given node depends significantly on the local field at that node.
\ Since the field at a unanimous bit is stronger than the field at a
non-unanimous bit, we expect the magnitude of $x_{i}$ to be larger for a
unanimous bit. \ Therefore, we make the \textit{ansatz}%
\begin{equation}
x_{i}^{S}=\left\{
\begin{array}
[c]{ll}%
A\operatorname{sign}(\xi_{i}^{\mu_{1}}+\xi_{i}^{\mu_{2}}+\xi_{i}^{\mu_{3}}) &
\text{if \ }\xi_{i}^{\mu_{1}}=\xi_{i}^{\mu_{2}}=\xi_{i}^{\mu_{3}}\\
D\operatorname{sign}(\xi_{i}^{\mu_{1}}+\xi_{i}^{\mu_{2}}+\xi_{i}^{\mu_{3}}) &
\text{otherwise}%
\end{array}
\right.  \label{BGNthreemix}%
\end{equation}
where $A$ and $D$ are real numbers. \ The dynamical equations for the network
give a pair of self-consistency equations which can be solved numerically for
$A$ and $D.$ \ First, we need an expression for the overlap of the mixture
state with one of the three patterns, say, pattern $\mu_{1}$. If the $i$-th
bit is a unanimous bit, then $x_{i}^{S}$ has magnitude $A$ and agrees in sign
with $\xi_{i}^{\mu_{1}}$. On the other hand, if it is a non-unanimous bit,
then $x_{i}^{S}$ has magnitude $D$ and has a $2/3$ probablity of agreeing in
sign with $\xi_{i}^{\mu_{1}}$. The result is that%

\begin{equation}
m_{\mu_{1}}=\frac{1}{N}\sum\limits_{i}x_{i}\xi_{i}^{\mu_{1}}=\frac{1}%
{4}A+\frac{1}{2}D+\frac{1}{4}(-D)=\frac{1}{4}(A+D). \label{BGNthreeOv}%
\end{equation}
Note that in the special case $A=D=1$, the above expression reduces to the
Hopfield value $\ 1/2$ , as it should. Again, all three overlaps have the same
size: $m_{\mu_{1}}=m_{\mu_{2}}=m_{\mu_{3}}$\ . \ The total energy of the
network in this state is given by%
\begin{align}
E(A,D)  &  =\sum\limits_{i}(\frac{x_{i}^{4}}{4}-\frac{x_{i}^{2}}{2})-\frac
{N}{2}\gamma\sum\limits_{\mu}m_{\mu}^{2}\nonumber\\
&  =\frac{N}{4}\left(  \frac{A^{4}}{4}-\frac{A^{2}}{2}\right)  +\frac{3N}%
{4}\left(  \frac{D^{4}}{4}-\frac{D^{2}}{2}\right)  -\frac{3N}{2}\gamma\left(
\frac{1}{4}(A+D)\right)  ^{2}\nonumber\\
&  =N\left(  \frac{A^{4}}{16}+\frac{3D^{4}}{16}-\frac{A^{4}}{8}-\frac{3D^{2}%
}{8}-\gamma\left(  \frac{3A^{2}}{32}+\frac{3D^{2}}{32}+\frac{3AD}{16}\right)
\right)  . \label{BGNthreeE}%
\end{align}
We can think of this as an an energy function on the restricted family of
states parametrized by (\ref{BGNthreemix}). \ 

\begin{figure}
[ptb]
\begin{center}
\includegraphics[
height=3.947in,
width=3.0952in
]%
{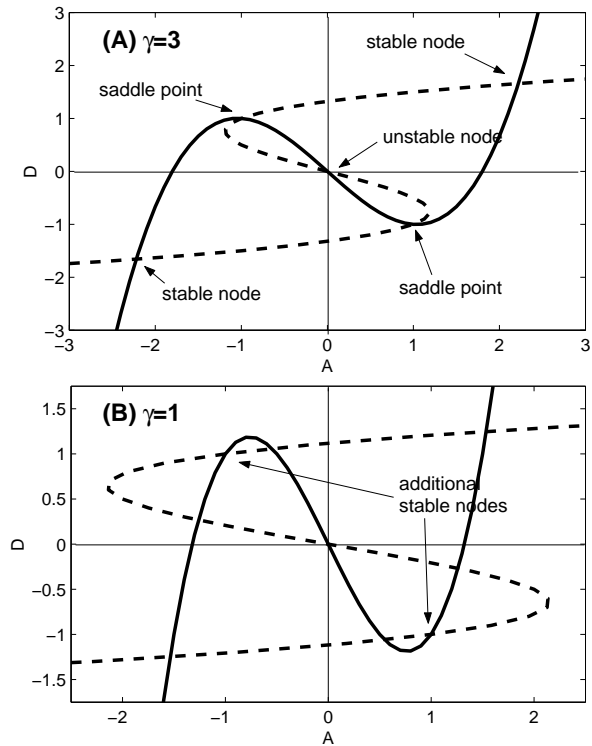}%
\caption{Self-consistency equations for the $n=3$ mixture state. (A) \ Stable
solutions occur in the first and third quadrants. \ (B) \ When $\gamma<2$ two
additional stable solutions appear in the second and fourth quadrants. \ }%
\label{nullclinesboth}%
\end{center}
\end{figure}

A necessary condition for the mixture state (\ref{BGNthreemix}) to be a fixed
point is that $\frac{\partial E}{\partial A}=\frac{\partial E}{\partial D}=0$.
This gives us two self-consistency conditions for the parameters $A$ and $D$:
\begin{align}
\frac{A^{3}}{4}-\frac{A}{4}-\frac{3\gamma}{16}(A+D)  &  =0\nonumber\\
\frac{3D^{3}}{4}-\frac{3D}{4}-\frac{3\gamma}{16}(A+D)  &  =0.
\label{selfconsist}%
\end{align}
These are the equations of \ the nullclines of the energy function
(\ref{BGNthreeE}). \ Alternatively, the above equations could be derived
directly from the dynamical equations for each node and the expressions for
$h_{i}$ instead of using the energy function. \ For a \emph{stable} fixed
point, $(A,D)$ must be a local minimum of the energy function (\ref{BGNthreeE}%
). The self-consistency equations (\ref{selfconsist}) can be rewritten as
follows:%
\begin{align}
D  &  =\frac{4}{3\gamma}\left[  A^{3}-(1+\frac{3\gamma}{4})A\right]
\nonumber\\
A  &  =\frac{4}{\gamma}\left[  D^{3}-(1+\frac{\gamma}{4})D\right]  .
\end{align}
The graphs of these two cubic equations are plotted for $\gamma=3$ in figure
\ref{nullclinesboth}A. \ Solutions are points where the two nullclines
intersect. \ \ Note that the slope, $dD/dA$ , of the first curve at the origin
is $-\left(  1+\frac{4}{3\gamma}\right)  <-1,$ and for the second curve
$\frac{dD}{dA}=-\frac{1}{1+\frac{4}{\gamma}}>-1$ . \ These two inequalities
satisfied by the slopes ensure that the curves always intersect in at least 5
points. \ The five solutions can be classified by looking at the energy
function and its gradient. We see that the solution $A=D=0$ is an unstable
node (maximum of the energy), the two solutions in the 2nd and 4th quadrants
are saddle points, and the two solutions in the first and third quadrants are
the stable solutions we seek. (There are two because of the $Z_{2}$ sign
reversal symmetry --- one is a mirror state of the other.)%
\ 

\begin{figure}
[ptb]
\begin{center}
\includegraphics[
height=3.8614in,
width=4.7392in
]%
{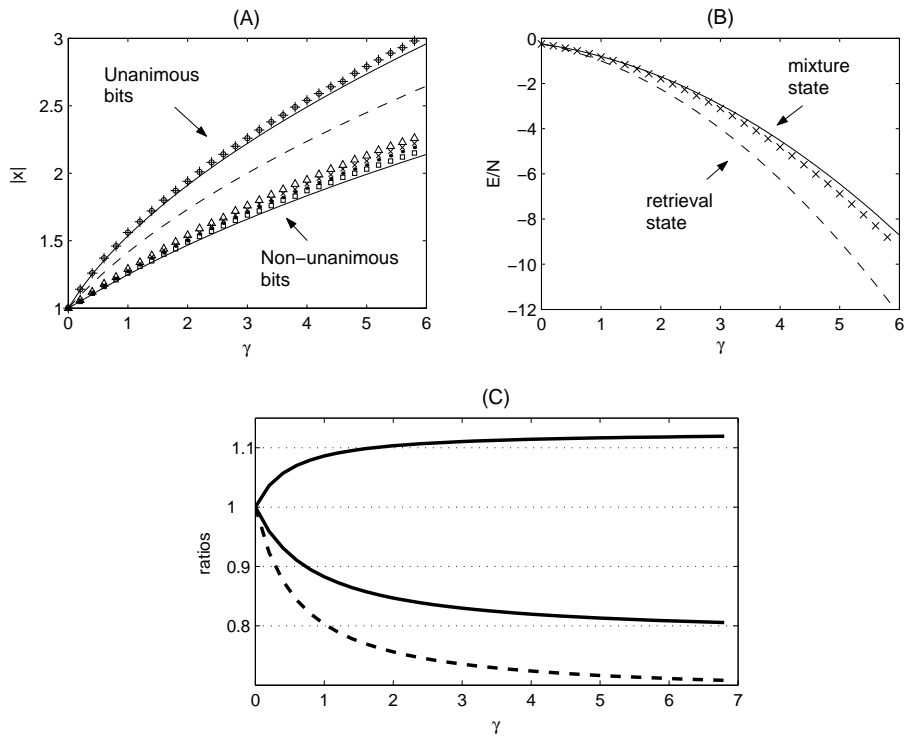}%
\caption{n=3 mixture state compared with retrieval state. \ (A) \ The two
solid lines represent A and D, the values of $\left\vert x\right\vert $ for
unanimous and non-unanimous bits respectively, obtained by numerical solution
of the self-consistency equations. The symbols show the \emph{observed} values
$\left\vert x_{i}\right\vert \;(1\leq i\leq6)$ for a mixture state of the
dynamically simulated network. \ The dotted line is $\sqrt{1+\gamma}$, which
is the theoretical value of all $\left\vert x_{i}\right\vert $ for a retrieval
state. \ (B) Energy $E_{mix}$ for the mixture state. Solid line=solution of
self-consistency equations; symbols=observed energy of simulated mixture
state; dotted line=energy $E_{ret}$ of retrieval state from eq.
(\ref{retrievalE}). \ (C) the ratios $A/\sqrt{1+\gamma}$ and \ $D/\sqrt
{1+\gamma}$ (upper and lower solid lines) and $E_{mix}/E_{ret}$ (dotted line)
approach asymptotically constant values. \ }%
\label{mixcompare}%
\end{center}
\end{figure}

The self-consistency equations for $A$ and $D$ were solved numerically for a
range of $\gamma$ values using a gradient descent algorithm. \ These values
were in turn substituted into (\ref{BGNthreeE}) to find the energy as a
function of $\gamma$. The results are plotted in figure \ref{mixcompare} where
they are also compared with numerical results from dynamical simulations of a
BGN with $N=1000$. We studied the mixture state numerically by initializing a
network to the state $x_{i}=\operatorname{sign}(\xi_{i}^{1}+\xi_{i}^{2}%
+\xi_{i}^{3})$ and incrementing $\gamma$ beginning at 0, much as was done for
the retrieval states. \ Figure \ref{mixcompare}A shows the magnitudes of
\ unanimous and non-unanimous bits in the mixture state as functions of
$\gamma.$ The solid lines show the solutions of the self consistency equations
for $A$ and $D$. \ The symbols show the observed magnitudes $\left\vert
x_{i}\right\vert $ for $1\leq i\leq6$ in the simulated network state. \ Two of
these first six bits are unanimous while the other four are not. \ There is
good agreement between the observed values of $\left\vert x_{i}\right\vert
$and the values obtained from the self-consistency equations. \ For
comparison, $\sqrt{1+\gamma}$ is plotted as a dotted line on the same axes.
\ (Recall that this is the value of all $\left\vert x_{i}\right\vert $ in a
pure retrieval state.) \ \ Figure \ref{mixcompare}B shows a corresponding
comparison of the observed and theoretical energies. \ Finally, figure
\ref{mixcompare}C \ compares the $n=3$ mixture state with the retrieval state
by plotting the ratios \ $A/\sqrt{1+\gamma}$ and $D/\sqrt{1+\gamma}$ as well
as the ratio of the mixture state energy $E_{mix\text{ }}$to that of a
retrieval state $E_{ret}$. \ All three of these ratios appear to approach
constant asymptotic values as $\gamma$ increases. \ Asymptotically,
\ $E_{mix}/E_{ret}\approx0.7$, \ while for the HN the corresponding ratio is
0.75. \ \ \ The strength of the field acting on each unanimous bit, $h_{A}$,
\ and that acting on the non-unanimous bits, $h_{D}$, both increase as
$\gamma$ increases. \ \ The mixture state is stable against single sign flips
of the unanimous bits when $h_{A}>\frac{2\sqrt{3}}{9}$ and stable against any
single sign flip when $h_{D}>\frac{2\sqrt{3}}{9}$. \ Thus as $\gamma$
increases, the mixture state begins to develop a non-trivial basin of
attraction of its own.%

\ \ 

It is interesting to note that at $\gamma=2$ a saddle-node bifurcation occurs
and for $\gamma<2$ two aditional stable solutions to the self-consistency
equations (\ref{selfconsist}) appear in the second and fourth quadrants, at
$(A,D)=(1,-1)$ and $(-1,1)$. \ (See figure \ref{nullclinesboth}B.) \ These are
states with $E=-\frac{1}{4}N$ \ in which $m_{\mu_{1}}=m_{\mu_{2}}=m_{\mu_{3}%
}=0$ \ and the net field acting on each node exactly cancels. \ They are not
stable against sign flips, and become completely destabilized when $\gamma>2$.
\ As we will see in the following section, they do not properly belong to the
category of mixture or spin glass states, but rather to another class of
spurious attractors present only in the BGN at low values of $\gamma$. \ 

Here we examined only the $n=3$ mixture state, \ but similar methods may be
used to characterize higher-order mixtures. \ In general, they are more
complex as there are more possibilities for the size of the majority by which
the sign is determined. \ The magnitudes of $x_{i}$ then take a greater number
of distinct values. \ 

\section{Uncondensed states and their collapse to the pattern subspace}

In section \ref{errorcorrect}, we noted that for values of $\gamma$ not far
above $\frac{1}{3}$, \ it is possible that a state may have a significant
overlap with one of the stored patterns but that the field acting on the nodes
may nonetheless not be strong enough to overcome the potential barrier and
correct the sign errors. \ Indeed if $\gamma$ is below $\frac{1}{3},$ then
even a single sign error may go uncorrected.\ This consequence of the
bistability of the BGN units contrasts with the behavior of the HN. \ 

Consider first the case of the HN. A typical random initial state has small
but nonzero overlaps with the memorized patterns, \ $m_{\mu}\sim O(1/\sqrt
{N})$, resulting in fields $h_{i}=\sum\xi_{i}^{\mu}m_{\mu}$ which are random
with zero mean and variance of order $1/\sqrt{N}$. \ \ Typically, for
approximately half of the nodes $x_{i}$ and $h_{i}$ will initially have
opposite signs. \ Since there is no potential barrier against sign flips,
those nodes will change their signs, and the sign flips will continue until
the field experienced by every node is aligned with $x_{i}$. \ Every sign flip
will increase the magnitude of one or more of the overlap variables. \ If, for
example, \ one overlap $m_{\nu}$ is larger than all of the others, then most
nodes will experience fields which tend to align them with pattern $\nu$.
\ Every sign flip further increases the value of $m_{\nu}$, and eventually
$m_{\nu}$ will be fully retrieved even though the initial overlap may have
been quite small. \ However, if one overlap does not clearly dominate the
others, then the trajectory may arrive at a spurious attractor which has
roughly equal overlaps with several patterns, instead of at a single one of
the memory states. \ Even a state which is initially orthogonal to all of the
memorized patterns can be rendered unstable by changing a single sign: \ even
a single sign flip will create a small but nonzero field affecting the other
nodes, resulting in further sign flips, and so on. \ In summary, for the HN
essentially any initial condition converges under the dynamics to an attractor
lying in or close to the $p$-dimensional subspace of the patterns.

For the BGN, on the other hand, the situation is different due to the presence
of potential barriers. \ Just as with the HN, given any state $\mathbf{x}$
every node $i$ experiences a field $h_{i}$, which may be aligned with or
opposed to $x_{i}$. \ However, the antiparallel local fields may not be strong
enough to flip their nodes into the parallel direction. \ \ If most $h_{i}$
are well below the threshold $\frac{2\sqrt{3}}{9}$ then the flipping of one or
a few nodes will not change the field enough to cause any further flips.
\ \ Thus there might be a large number of initial conditions which remain
stuck with low overlaps, far away from any of the patterns. \ We refer to such
states with sub-threshold fields as \textquotedblleft
uncondensed\textquotedblright\ states, because in those states none of the
order paramenters $m_{\mu}$ are condensed. \ However, we will show below that
states with low overlaps cannot remain stable for $\gamma>2$, \ and thus for
higher values of $\gamma$ the behavior is Hopfield-like, with all trajectories
collapsing toward the pattern subspace. \ 

Consider a hypothetical state which is strictly orthogonal to all memory
patterns, so that $m_{\mu}=0$ for all $\mu$. \ (The extra solutions appearing
in the self-consistency equations for the mixture state when $\gamma<2$ are
examples of such states.) If $m_{\mu}=0$ for all $\mu$, then $h_{i}=0$ for all
$i$. \ In this case, the steady state of each node is $x_{i}=\pm1$.
\ Proceeding with linear stability analysis as above, we find the relevant
Jacobian%
\begin{equation}
\frac{\partial y_{i}}{\partial x_{j}}=(1-3x_{i}^{2})\delta_{ij}+\gamma
w_{ij}=-2\delta_{ij}+\gamma w_{ij}.
\end{equation}
The equilibrium is unstable if $\gamma w_{max}>2$ , where $w_{\max}$ is the
largest eigenvalue of the coupling matrix. $w_{\mathrm{\max}}$ is at least 1.
Therefore, if $\gamma>2$, states orthogonal to the stored patterns are all
unstable. \ If all of the $p$ stored patterns are mutually orthogonal (which
is approximately true in the limit $N\rightarrow\infty$) \ then for $\gamma>2$
there are $p$ unstable eigenvalues. \ 

We can examine one of those unstable directions, say, the one associated with
the $\nu$-th pattern, more closely by making an explicit \emph{ansatz.}\ Let
us denote the orthogonal, zero-overlap state by $x_{i}=\xi_{i}^{\bot}$; by
assumption $m_{\mu}=\sum_{i}\xi_{i}^{\bot}\xi_{i}^{\mu}=0$ for all $\mu$
including $\mu=\nu$. \ For half of the nodes $\xi_{i}^{\bot}=\xi_{i}^{\nu}$;
for the other half $\xi_{i}^{\bot}=-\xi_{i}^{\nu}$. \ Let us then consider a
family of states described by the \emph{ansatz}%
\begin{equation}
x=\left\{
\begin{array}
[c]{ll}%
A\xi_{i}^{\bot}=A\xi_{i}^{\nu} & \text{if }\xi_{i}^{\bot}=\xi_{i}^{\nu}\\
B\xi_{i}^{\bot}=-B\xi_{i}^{\nu} & \text{if }\xi_{i}^{\bot}=-\xi_{i}^{\nu}%
\end{array}
\right.  \label{unstable}%
\end{equation}
This parametrizes a 2-dimensional subspace of the state space which contains
$\mathbf{\xi}^{\bot}$ and one of its unstable eigenvectors. \ Using the
methods of section \ref{mixture}, \ one obtains a pair of cubic
self-consistency equations for $A$ and $B$:%
\begin{equation}
B=-\frac{A^{3}}{\gamma}+\left(  \frac{1}{\gamma}+1\right)  A,\;\;A=-\frac
{B^{3}}{\gamma}+\left(  \frac{1}{\gamma}+1\right)  B.
\end{equation}
\ \ For all $\gamma$ there are two solutions $(A,B)=(\pm\sqrt{1+\gamma}%
,\mp\sqrt{1+\gamma}),$ which correspond to the retrieval states $\mathbf{x}%
=\pm\sqrt{1+\gamma}\mathbf{\xi}^{\nu}$. \ For $\gamma<2$ there are additional
stable solutions $(A,B)=\pm(1,1)$ corresponding to $\pm$ $\mathbf{\xi}^{\bot}%
$. \ A bifurcation occurs at $\gamma=2$ and these solutions become saddle
points. \ 

Thus, \ we see that there is an absolute upper limit for the existence of
stable uncondensed states. \ \ \ In fact, $\gamma=2$ turns out to be a high
upper bound. \ \ The example of a state with all $m_{\mu}$ equal to zero is a
sort of \textquotedblleft worst-case scenario.\textquotedblright\ \ For a
finite-sized network the typical random initial condition has small but
nonzero overlaps. \ In addition, if the patterns are truly random then they
will not be exactly orthogonal but have small overlaps and so the largest
eigenvalue of the synaptic matrix will be slightly larger than unity.
\ Because of these factors the typical uncondensed state becomes unstable at
values of $\gamma$ lower than 2; in \ numerical simulations we found that for
the case $N=1000,\quad p=5$ most become unstable between $\gamma=1$ and $1.5$.\ \ 

\begin{figure}
[ptb]
\begin{center}
\includegraphics[
height=2.821in,
width=3.8112in
]%
{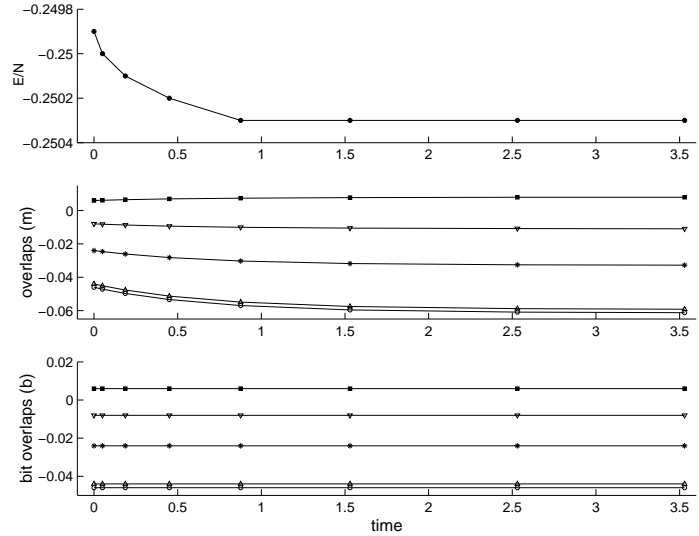}%
\caption{Trajectory of a random initial condition for $\gamma=0.5$. \ Note:
\ The unequal time steps result from the adaptive step size control in our
integration algorithm.}%
\label{dyna05}%
\end{center}
\end{figure}

\begin{figure}
[ptb]
\begin{center}
\includegraphics[
height=2.9793in,
width=3.9263in
]%
{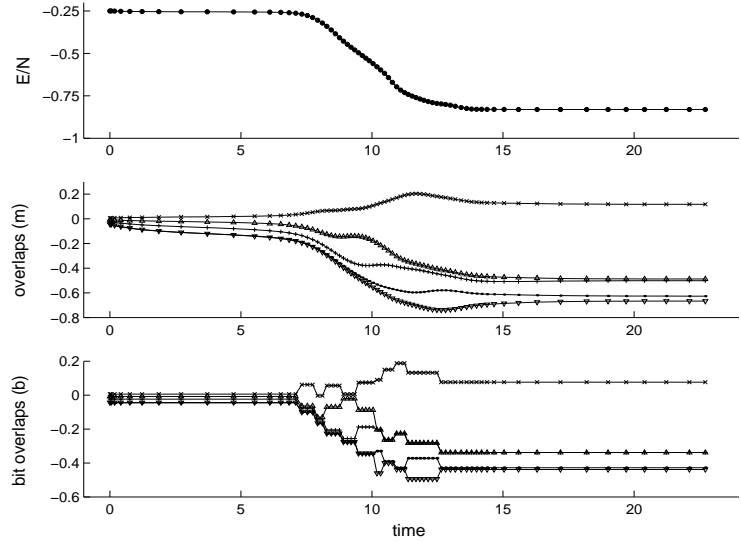}%
\caption{Trajectory of the same initial condition, $\gamma=1.0$. }%
\label{dyna10}%
\end{center}
\end{figure}

\begin{figure}
[ptbptb]
\begin{center}
\includegraphics[
height=4.0015in,
width=3.8415in
]%
{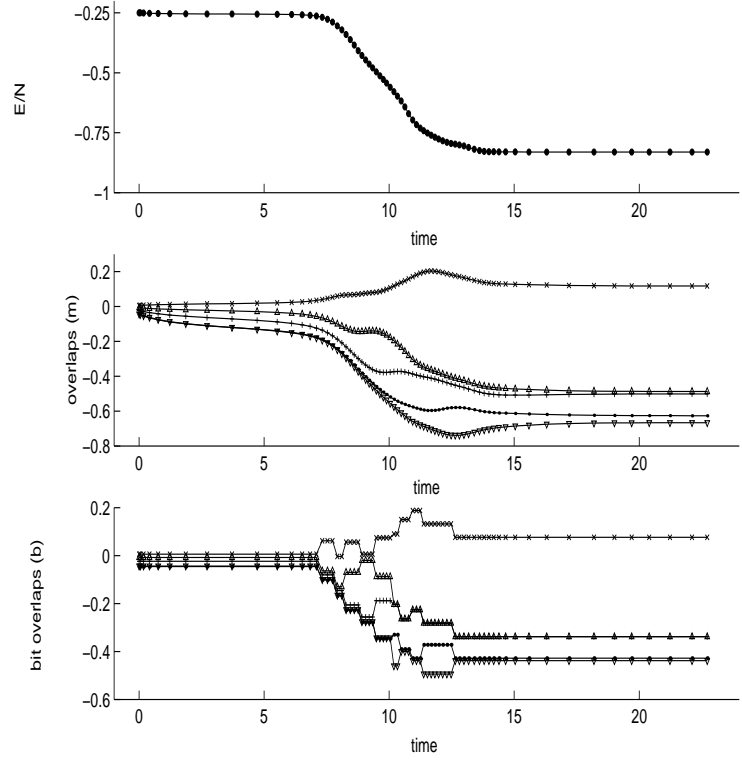}%
\caption{ Same initial condition, $\gamma=1.2.$ }%
\label{dyna12}%
\end{center}
\end{figure}

\begin{figure}
[ptbptbptb]
\begin{center}
\includegraphics[
height=4.0663in,
width=3.8173in
]%
{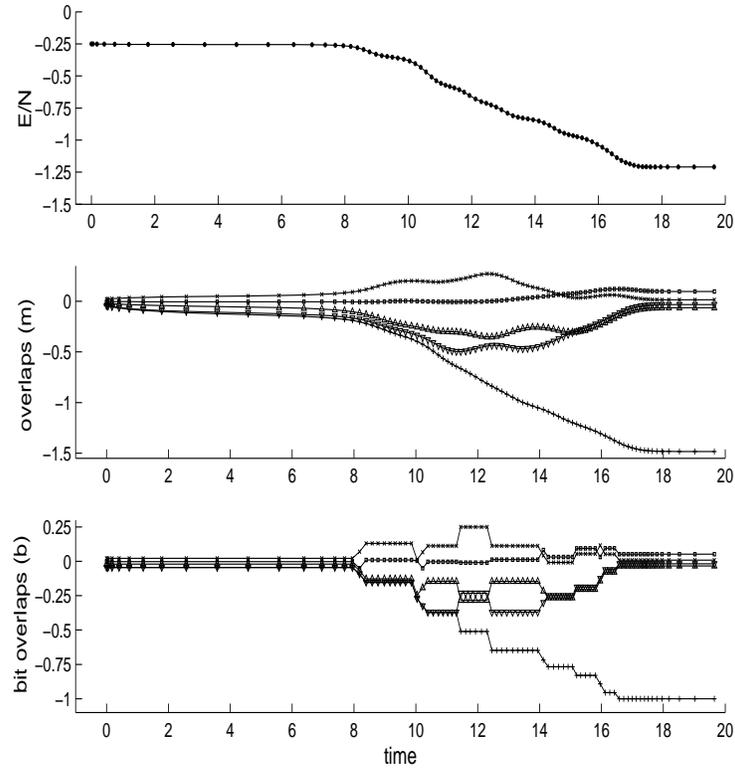}%
\caption{An $\gamma=1.2$ trajectory with a different random initial condition.
\ In this case, one of the bit overlaps reaches $-1$, while the others become
small. \ This indicates that the mirror state of one of the stored patterns
has been retrieved. \ }%
\label{dyn12b}%
\end{center}
\end{figure}

Figures \ref{dyna05}-\ref{dyn12b}, we show numerical results for the fate of a
typical random initial condition of the BGN with $N=1000$, $p=5$. \ Figures
\ref{dyna05}-\ref{dyna12} show the dynamical evolution of the same initial
condition, at different values of $\gamma$. \ The initial condition was a
random string of $\pm1$ values. \ We plot the energy per node, all five
overlap variables $m_{\mu}$ and all five bit overlap variables $b_{\mu}$ as
functions of time. \ Recall that $b_{\mu}$ contain information about sign
agreements only. \ For $\gamma=0.5$ (figure \ref{dyna05}) the state changes
very little before convergence occurs. \ The energy per node remains very
close to $-0.25$. The overlaps $m_{\mu}$ increase slightly in
magnitude\footnote{Note that negative overlaps with a pattern can be viewed as
positive overlaps with the corresponding mirror state.}, but the bit overlaps
do not change at all, indicating that no sign flips occur. \ For $\gamma=1$
(figure \ref{dyna10}) \ the trajectory is similar, except that the small
initial overlaps are amplified to a greater extent (we will explain this
effect below.) \ The bit overlaps still do not change. \ When $\gamma=1.2$,
\ however, the trajectory changes qualitatively (figure \ref{dyna12}). The
magnitudes of the overlaps $m_{\mu}$ grow slowly until at $t\sim7$ the
resulting field becomes strong enough to begin flipping some signs. At this
point the bit overlaps begin to change, the energy drops significantly and the
trajectory moves close to the pattern subspace. \ \ After some further
evolution, the system converges to a mixture state which overlaps with several
patterns. \ A different random initial condition, followed again at
$\gamma=1.2$, leads instead to a memory state (figure \ref{dyn12b}). \ Here
one of the five overlaps becomes dominant and the others shrink away. \ In
this case the mirror state of one of the five patterns is retrieved. \ These
trajectories are typical examples representing descent on a rugged energy
landscape. \ Different initial conditions lead to different attractors, of
which some are memory states and some are mixtures. \ Frequently the
trajectory lingers at one or several states before settling at its asymptotic
attractor. \ %

%


Figures \ref{dyna05}-\ref{dyna10} illustrated that for sufficiently small
values of $\gamma$, \ the dynamics amplifies small initial overlaps without
flipping the sign of any node. \ For more insight into this phenomenon,
consider an initial state in which all $x_{i}$ are $\pm1$ but somewhat more
nodes are aligned parallel with one particular pattern $\xi_{i}^{\nu}$ than
are antiparallel. \ In other words, $b_{\nu}$ is nonzero but less than unity.
\ \ For simplicity let us neglect all other overlaps. \ \ Initially, each node
experiences a small field given by $h_{i}=\gamma m_{\nu}\xi_{i}^{\nu}=\gamma
b_{\nu}\xi_{i}^{\nu}.$ \ This field will push $x_{i}$ to larger magnitudes
$(>1)$ for those nodes which are aligned with pattern $\xi_{i}^{\nu}$, \ and
it will push the others to smaller magnitudes $x_{i}<1$. \ This adjustment in
turn increases the value of $m_{\nu}$, until an equilibrium is reached with
$m_{\nu}>b_{\nu}$. \ We might think of this as a kind of \ \textquotedblleft
subliminal\textquotedblright\ recognition of the pattern. \ The effect becomes
stronger as $\gamma$ increases. \ Clearly it has a nonlinear dependence on
both $\gamma$ and $b_{\nu}$. \ When the field becomes large enough it will
exceed the threshold for sign flips and the state will be attracted toward the
pattern retrieval state. \ The larger $\gamma$ is, the smaller the initial
$b_{\nu}$ that is necessary to fully retrieve the pattern $\xi_{i}^{\nu}$.
\ \ In other words, the basins of attraction of the patterns expand as
$\gamma$ increases. \ 

\section{Basins of Attraction and the Energy Landscape}

In this section, we provide numerical support for the three-way classification
of attractors into retrieval, spin glass and uncondensed states and we show
how the respective attractor basins change with the control parameter $\gamma
$. \ We observe an interpolation between two different regimes. \ As we showed
above, for $\gamma\gtrsim2$ there are no stable uncondensed states. \ For
lower values of $\gamma$, on the other hand, uncondensed states are numerous.
\ Recall that uncondensed states are characterized by local fields too weak to
overcome the potential barriers against sign flips, \ and so their dynamics is
dominated by the local potential. \ In the extreme case $\gamma=0$, \ there
are of course no magnetic fields at all and only the local potential is present.

\subsection{Statistics of attractors reached from random initial conditions}

\begin{figure}
[ptb]
\begin{center}
\includegraphics[
height=4.0949in,
width=5.1906in
]%
{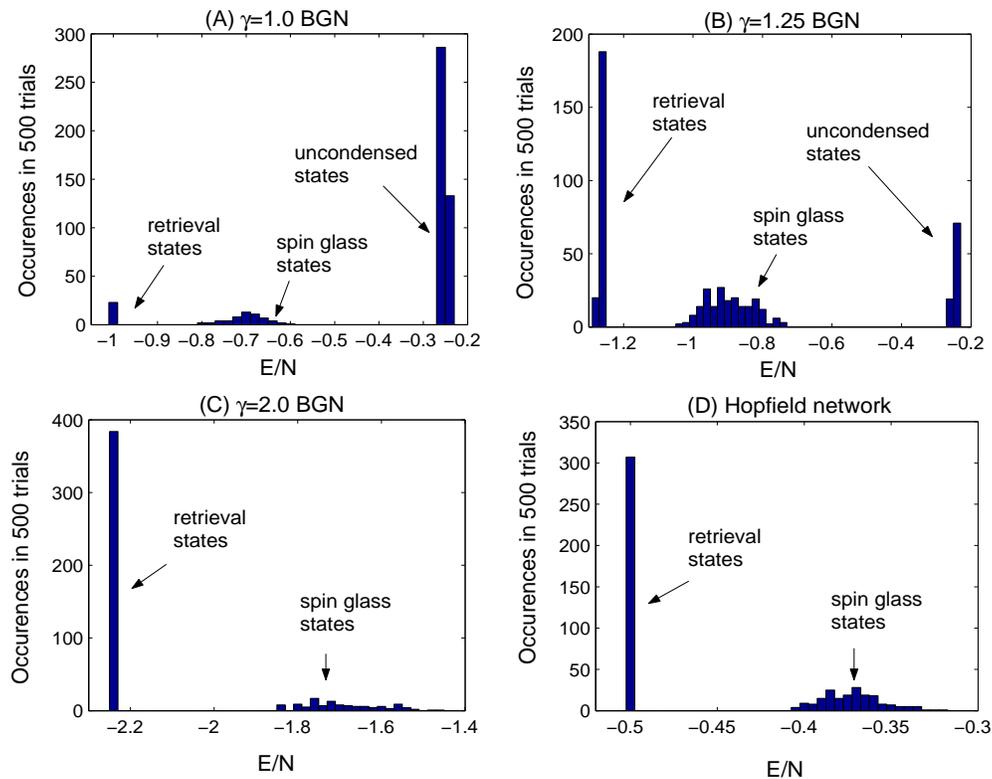}%
\caption{Energy per node for attractors reached from random initial
conditions, showing a clear separation among different types. \ (A) For
$\gamma=1$, three types of attractors are clearly present. \ Uncondensed
states show up as a peak near $E=-0.25N.$ (B) For $\gamma=1.25,$ the
uncondensed state peak is smaller but occurs at nearly the same energy,
whereas the other two peaks are at different energies.\ (C) For $\gamma=2$,
only the retrieval and spin glass states are obtained. \ (C) HN behaviour is
qualitatively similar to the BGN with $\gamma=2$. \ }%
\label{seedhist}%
\end{center}
\end{figure}

The classification of attractors is very clearly reflected in the energy
spectrum. \ To explore attractors and their basins, we \textquotedblleft
seeded\textquotedblright\ the network with 500 random initial conditions
(taken with several different realizations of the five random patterns),
integrated the dynamical equations until they converged, and constructed a
histogram of the final energies (figure \ref{seedhist}). \ For the case
$\gamma=1$ (figure \ref{seedhist}A), there are three clearly separated
clusters of attractors. \ Those with the lowest energies are retrieval states,
\ while the states clustered at $E/N=-0.25$ are the uncondensed states, and
those in the intermediate range are the glassy states. \ The picture is
qualitatively similar at the slightly larger value $\gamma=1.25$ (figure
\ref{seedhist}B), \ but the peak at $E/N=-0.25$ has shrunk relative to the
other two. \ Note that the energies of the retrieval and spin glass states
change with $\gamma$, \ while the uncondensed states remain at nearly the same
energy because their dynamics is dominated by the local potential. \ \ For
$\gamma=2$, (figure \ref{seedhist}C) \ on the other hand, the cluster at
$E/N=-0.25$ is absent as there are no stable uncondensed states. \ The
histogram for a HN (figure \ref{seedhist}D) resembles that for the BGN with
$\gamma=2$. \ \ One quantitative difference is that the retrieval state peak
is slightly higher for the BGN with $\gamma=2$, while the glassy states are
comparatively suppressed. \ 

\begin{figure}
[ptbptb]
\begin{center}
\includegraphics[
height=2.9793in,
width=3.7559in
]%
{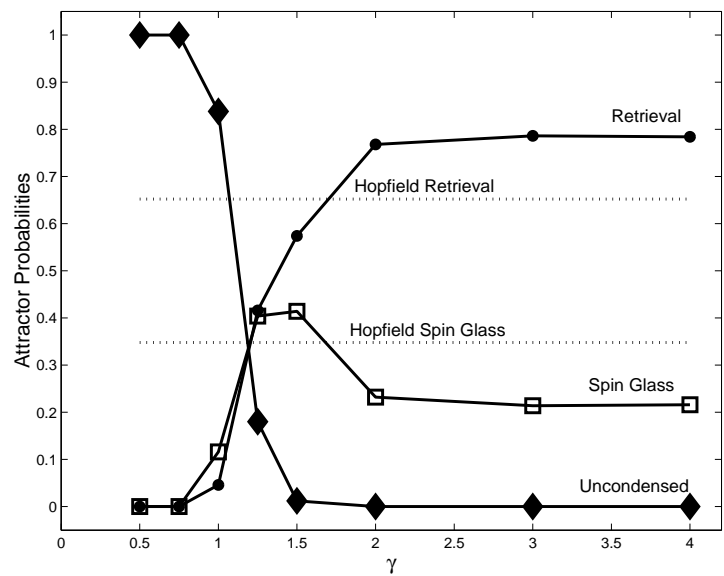}%
\caption{Probability of convergence of a random initial condition to each of
the three types of attractors, plotted as functions of $\gamma$. \ Hopfield
values are shown for comparison. \ }%
\label{attractorprobs}%
\end{center}
\end{figure}

We performed this experiment at a range of values of $\gamma$. \ In all cases
the classification of states was clear from the energy spectrum and was
verified by examining the final values of $b_{\mu}$. \ Figure
\ref{attractorprobs} shows the probabilities of convergence to each of the
three types of attractors from a random initial condition as functions of
$\gamma$. \ At $\gamma=0.5$ the landscape is dominated by the uncondensed
states. \ Even though $\gamma=0.5$ lies above the threshold of $\frac{1}{3}$
and the retrieval states have non-trivial basins of attraction, these basins
still occupy a very small fraction of the total configuration space volume.
\ The patterns can be retrieved only if the initial overlaps are relatively
high, and the probability of a \emph{random} initial condition being
sufficiently close is very low. \ The retrieval probability becomes
significant only as $\gamma$ approaches 1. \ As $\gamma$ increases from 1 to
1.5, basins for the memory and spin glass states grow at the expense of of the
uncondensed states until the latter disappear. \ The retrieval state basins
grow faster than those of the spin glass states. \ Beyond $\gamma=1.5$, \ the
probability of retrieving a memory state saturates at approximately a $10\%$
higher value than in the Hopfield case, and the probability of falling into a
spin glass state is correspondingly lower.%

\

\subsection{Mapping the boundaries of basins of attraction}

\begin{figure}
[ptb]
\begin{center}
\includegraphics[
height=4.0179in,
width=4.7418in
]%
{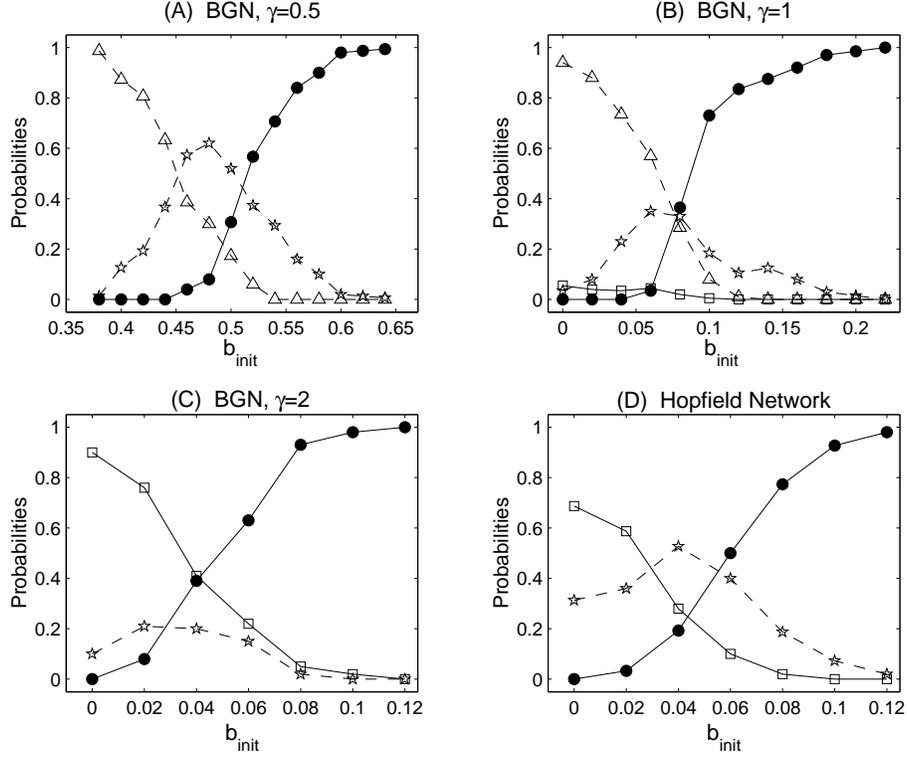}%
\caption{Attractors retrieved from states with a specified initial overlap
with a target pattern. \ Plots show probability of retrieving the the target
pattern (solid circles), \ one of the other memory patterns (squares), \ an
uncondensed state (triangles) or a spin glass spurious state (stars). \ }%
\label{boundaries}%
\end{center}
\end{figure}

In an attempt to map the attractor basins in more detail, we generated
configurations at specified initial Hamming distances from particular memory
patterns. \ This was done by starting with a pattern $\mathbf{\xi}^{\nu}$ and
flipping the signs of a specified number of randomly chosen bits. \ \ Using an
ensemble of such initial conditions, we measured the probability of retrieval
of the target pattern $\mathbf{\xi}^{\nu}$ as a function of the initial
distance from it. \ As a rule, the probability of recognizing the pattern is
high if only a few signs are flipped, but drops sharply if a certain threshold
Hamming distance is exceeded. \ We are interested in learning where this
threshold lies, and thus answering the question of how close an initial
condition must be to a pattern in order to be attracted to it. \ We are also
interested in the fate of states lying just outside the boundaries of a basin
of attraction.\ \ In other words, does the basin share a boundary with the
basins of other patterns, or only with spurious attractors? \ The results are
presented in figure \ref{boundaries} for a BGN with $N=1000$ and $p=5$, for
the three values $\gamma=0.5,1.0,$ and $2.0$, and also for the HN. \ In each
of these cases, we generated an ensemble of initial conditions at a particular
initial value of $b_{\nu}$ for some pattern. \ Each initial condition was
allowed to evolve under the dynamics and the resulting attractor was
classified as: \ A) \ the target pattern $\mathbf{\xi}^{\nu},$ B) \ one of the
other patterns $\mathbf{\xi}^{\mu}\;(\mu\neq\nu),$ \ C) A spin glass spurious
state or \ D) \ an uncondensed state. \ In this case, we classified a state as
uncondensed\ if no sign flips occured during the dynamical evolution. \ \ The
probabilities of each of these four outcomes were averaged over several
realizations of the random patterns and plotted as functions of the initial
bit overlap $b_{\nu}$. \

Consider first the HN data from figure \ref{boundaries}D. \ A pattern can
evidently be retrieved even if the initial overlap is fairly small; the
probability is close to unity if $b_{init}\gtrsim0.1$. \ If the target pattern
is not retrieved, then either a spurious attractor or one of the other
patterns may be retrieved. \ There is a range of $b_{init}$ over which all
three probabilities are significant, indicating that the basins for the memory
states border on each other as well as those of spurious states. \ \ For an
$N=1000$ network, \ the expected magnitude of the overlap of a random state
with any given one of the stored patterns is $1/\sqrt{N}\approx.03$, \ which
is not much smaller than the apparent threshold of $b_{init}\approx0.1.$
\ \ This is consistent with the view that for the HN, \ a pattern is likely to
be retrieved as long as the initial overlap with that pattern is significantly
larger than all of the other overlaps. \ \ The $\gamma=2$ BGN (figure
\ref{boundaries}C) \ shares the qualitative features of the HN. \ Note however
that the probability of becoming trapped in a spurious state is smaller for
the BGN, consistent with the results in figure \ref{attractorprobs}. \ \ 

A contrasting case is the $\gamma=0.5$ BGN (figure \ref{boundaries}A). \ In
this case retrieval of the target pattern requires an initial overlap of more
than 0.5. \ Although this represents a significant basin, it is highly
unlikely that a \emph{random} initial condition will have such a large
overlap, thus explaining why random initial conditions almost never flow to a
memory state. \ The basins of the memory states are bordered only by spurious
states, not by other memory states. \ Interestingly, the states which lie
adjacent to the basin of a memory pattern are not all uncondensed-- some are
spurious states of the mixture or spin glass type. \ Examination of the states
retrieved near the boundary shows that these are typically \emph{asymmetric}
mixture states with one large overlap and two or more smaller (but greater
than random) overlaps. \ Finally, for $\gamma=1$ (figure \ref{boundaries}B),
\ the basins of the memory state are almost as large as in the HN case, \ and
near the boundaries there is a small but nonzero probability of retrieving one
of the other memory patterns, indicating that the basins of different memory
states almost touch each other. \ 

\subsection{Qualitative picture of the energy landscape}

\begin{figure}
[ptb]
\begin{center}
\includegraphics[
height=2.9784in,
width=3.9781in
]%
{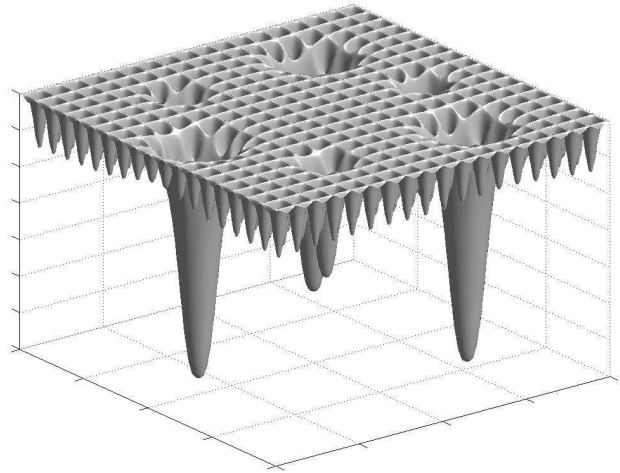}%
\caption{Schematic illustration showing qualitative features of the energy
landscape for low values of $\gamma$. \ The numerous small depressions
represent uncondensed states. \ The retrieval states and mixture states occupy
isolated valleys. \ (Retrieval states are represented by the deeper valleys.)
\ }%
\label{landscape05}%
\end{center}
\end{figure}

\begin{figure}
[ptbptb]
\begin{center}
\includegraphics[
height=2.9784in,
width=3.9781in
]%
{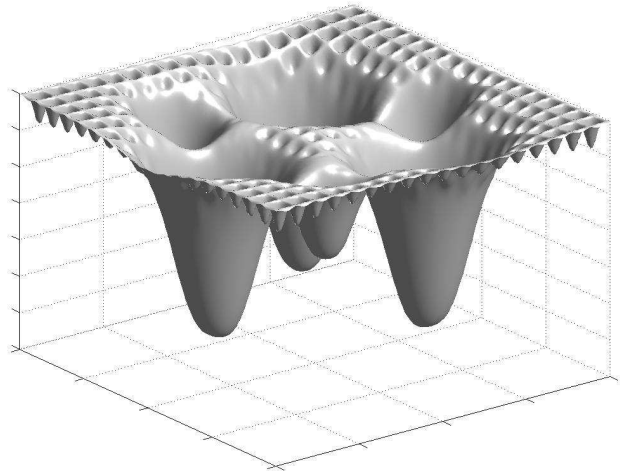}%
\caption{Energy landscape for intermediate values of $\gamma\sim1$. \ The
retrieval states and mixture states have large basins of attraction which
almost touch, but significant islands of uncondensed states remain. }%
\label{lanadscape1}%
\end{center}
\end{figure}

\begin{figure}
[ptbptbptb]
\begin{center}
\includegraphics[
height=2.9784in,
width=3.9781in
]%
{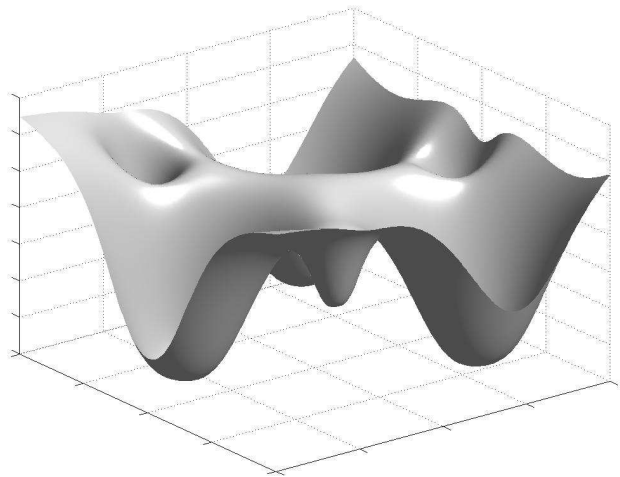}%
\caption{Energy landscape for large $\gamma$. \ There are no uncondensed
states, and large basins of attraction occupy the whole landscape. }%
\label{landscape2}%
\end{center}
\end{figure}

Taken together, the above results suggest a qualitative, schematic picture of
the energy landscape illustrated in figures \ref{landscape05}-\ref{landscape2}%
. \ The representation of the configuration space by two dimensions is not to
be taken literally, since it is of course $N$-dimensional. \ At low values of
$\gamma$, such as 0.5, the energy landscape is dominated by uncondensed
states, which form a series of shallow basins, each limited to roughly a
single orthant of $N$-dimensional space. \ These are represented in the
diagram by a series of shallow pits. \ The basins of attraction for the
retrieval and mixture states form isolated depressions in this pitted plateau.
\ \ They occupy nontrivial volumes but do not lie adjacent to each other (with
the exception of certain spin glass states which lie near the retrieval
states). At intermediate values $\gamma\approx1$, \ the basins of attraction
of the retrieval states are much larger and in some places almost touch each
other, \ but significant islands of uncondensed states remain. \ By $\gamma
=2$, \ however, the uncondensed states have disappeared and the basins of
attraction for the other two types of states occupy the entire energy
landscape and share boundaries with each other. \ %


\section{Conclusions}

We have studied the behaviour of the bistable gradient network in the
thermodynamic low-loading limits $N\rightarrow\infty$, $p/N\ll1$. \ We
described and classified the attractors of the dynamics and also observed the
effectiveness of pattern retrieval as a function of the coupling parameter
$\gamma$. \ We found that states corresponding to perfect retrieval of the
stored patterns are linearly stable at all values of $\gamma$, and have an
energy that decreases monotonically with $\gamma$. \ Above the threshold
$\gamma=\frac{1}{3}$ the retrieval states become stable against sign flips of
one or more nodes, and the network begins to function as an associative
memory. \ If $\gamma$ is not far above this threshold, then the basins of
attraction of the retrieval states are small and input must be very close in
Hamming distance to a pattern for recognition to occur. \ The basins of
attraction of the retrieval states grow as $\gamma$ increases. \ 

There are two regimes of behaviour, distinguished by the types of attractors
that occur. \ At low $\gamma$ the configuration space is dominated by the
uncondensed states, or states in which no node experiences a field strong
enough to overcome its potential barrier. \ In these states, $\left\vert
x_{i}\right\vert $ remains close to 1 for all nodes, \ and the energy remains
close to $-0.25$. \ Each of these states occupies a basin of attraction
confined to approximately a single orthant. \ In the limit $\gamma=0$ there
are $2^{N}$ such states, all degenerate in energy. \ As $\gamma$ increases
above the threshold $\gamma_{c}=\frac{1}{3}$, the retrieval states and the
mixture or spin glass states at first occupy small isolated basins among the
many uncondensed states. \ However, \ as $\gamma$ increases further, these
basins grow until they lie adjacent to each other. \ \ At some value of
$\gamma$ (observed to lie between 1 and 1.5), the uncondensed states disappear
and there is a transition to a Hopfield-like regime where the basins of
attraction for retrieval \ and spin glass states cover the whole configuration
space. \ As $\gamma$ increases still further, \ the retrieval basins grow at
the expense of the spin glass states, so that the latter can be noticeably
suppressed compared to the deterministic Hopfield case. \ \ This suppression
of the spurious states occurs without thermal noise or a modification of the
Hebb learning rule. \ 

The uncondensed states represent a phase which is neither ``ferromagnetic''
(i,e, strongly ordered and correlated with one pattern) nor ``glassy'' in the
sense that frustration is an important effect, \ yet they cannot properly be
described as ``paramagnetic,'' \ as paramagnetism is characterized by spins
which are able to flip freely from one orientation to the other. \ 

A few words on the application of such networks to practical problems of
associative memory are in order. \ The goal of associative memory is to
reconstruct a pattern from a more or less corrupted version or from a fragment
of the pattern, without becoming trapped in a spurious local minimum. \ From
this point of view, it appears that increasing $\gamma$ improves the
performance of the network--- expanding the basins of attraction for the
retrieval states and suppressing the spurious states. \ The low-$\gamma$
regime, on the other hand, may be suited to applications where the goal is a
\emph{selective} associative memory, \ one which only recognizes a pattern
from a fairly close approximation and thus avoids false recognition. \ In the
low-$\gamma$ regime, if the input is not close to one of the stored patterns,
then the network is likely to remain in an uncondensed state. \ These can in
general be distinguished clearly from other states (especially retrieval
states) by their relatively high energy ($E/N\approx-0.25$) \ or by the fact
that the magnitudes of all $\left\vert x_{i}\right\vert $ remain close to 1.
\ The magnitudes of the outputs can therefore be read as a signal of whether
recognition has occurred. \ Persistence in an uncondensed state corresponds to
an \textquotedblleft I don't know\textquotedblright\ or nonrecognition
response. \ 

In a subsequent publication, we will examine the behavior of the BGN when the
loading level $p/N$ is of order unity, and we will demonstrate another
performance trade-off. \ Specifically, we will show that the maximum storage
capacity of the network decreases as $\gamma$ increases. \ For a low-$\gamma$
regime, it is possible to stabilize more memorized patterns than in the
Hopfield case, while at higher $\gamma$, even though the low-loading fault
tolerance is increased, the storage capacity decreases. \ 

\begin{acknowledgments}
\bigskip This work was supported by Materials and Manufacturing Ontario (MMO),
a provincial centre of excellence.
\end{acknowledgments}

\bigskip

\bigskip

\bigskip

\bigskip

\bigskip

\bigskip

\bigskip

\bigskip

\bigskip\


\begin{thebibliography}{99}                                                                                               %


\bibitem {Haykin}S. Haykin, Neural Networks: A Comprehensive Foundation.
\ Prentice Hall, Upper Saddle River, NJ, 1999. \ 

\bibitem {Amitbook}Daniel J. Amit, \ Modeling Brain Function: \ The World Of
Attractor Neural Networks. \ Cambridge University Press 1989. \ 

\bibitem {Hopfield}J.J. Hopfield, \ Proc. Natl. Acad. Sci. USA \textbf{79},
\ 2554 (1982).

\bibitem {Little}W.A. Little, Math. Biosci. \textbf{19}, 101 (1974). \ W.A.
Little and G.L. Shaw, Math. Biosci. \textbf{39}, 281 (1978).

\bibitem {SKModel}S. Kirkpatrick and D. Sherrington, \ Phys. Rev.
\textbf{B17}, 4384 (1978).

\bibitem {Mezard}M. Mezard, G. Parisi and M. Virasoro, Spin Glass Theory and
Beyond. \ World Scientific, Singapore, 1987.

\bibitem {Chinarov}V. Chinarov and M. Menzinger, \ Biosystems \textbf{55}, 137
(2000). \ 

\bibitem {Hoppensteadt}F. C. Hoppensteadt and E. M. Izhikievich, Weakly
Connected Neural Networks, Springer-Verlag, New York 1997.

\bibitem {Bistability}T. Poston and I. Stewart, Behavioral Science
\textbf{23}, 318 (1978). \ \ I.N. Stewart and P.L. Peregoy, Physchological
Bulletin 94, 336 (1983). \ 

\bibitem {ContinuousHop}J.J. Hopfield, Proc. Natl. Acad. Sci. USA \textbf{81},
3088 (1984).

\bibitem {ChemNetE}W. Hohmann, M. Krauss and F.W. Schneider, J. Phys. Chem. A
\textbf{103}, 7606 (1999); \ J Phys. Chem A 102, 3103 (1998); \ J. Phys. Chem
A 101, 7364 (1997). \ G. Dechert, K.-P. Zeyer, D. Lebender and F.W. Schneider,
J. Phys Chem A 100, 19043 (1996).

\bibitem {ChemNetM}J.-P. Laplante, M. Pemberton, A. Hjelmfelt and J. Ross, J.
Phys. Chem 99, 10063 (1995). \ V. Booth, T. Erneux and J.-P. Laplante, J.
Phys. Chem 98, 6537 (1994). \ A. Hjelmfelt and J. Ross, J. Phys. Chem. 97,
7998 (1993). \ 

\bibitem {Amit1}D.J. Amit, H. Gutfreund and H. Sompolinsky, Phys.Rev.
\textbf{A32}, 1007 (1985); Phys. Rev. Lett. \textbf{55}, 1530 (1985); \ Ann.
Phys. \textbf{173}, 30 (1987). \ A. Chrisanti, D.J. Amit and H. Gutfreund,
Europhys. Lett. \textbf{2}, 337 (1986).

\bibitem {Hebb}D.O. Hebb, The Organization of Behavior, Wiley, New York, 1949. \ 
\end{thebibliography}
\end{document}